\documentclass[aps,pra,twocolumn,nofootinbib]{revtex4-1}

\usepackage{amssymb}
\usepackage{graphicx}
\usepackage{amsmath}
\usepackage{braket}
\usepackage{hyperref}
\usepackage{bbold}

\begin{document}

\title{Reexamination of Pure Qubit Work Extraction}

\author{Max F. Frenzel}
\affiliation{Controlled Quantum Dynamics Theory Group, Imperial College London, Prince Consort Road, London SW7 2BW, UK}
\author{David Jennings}
\affiliation{Controlled Quantum Dynamics Theory Group, Imperial College London, Prince Consort Road, London SW7 2BW, UK}
\author{Terry Rudolph}
\affiliation{Controlled Quantum Dynamics Theory Group, Imperial College London, Prince Consort Road, London SW7 2BW, UK}
\date{\today}

\begin{abstract}
Many work extraction or information erasure processes in the literature involve the raising and lowering of energy levels via external fields. But even if the actual system is treated quantum mechanically, the field is assumed to be classical and of infinite strength, hence not developing any correlations with the system or experiencing back-actions. We extend these considerations to a fully quantum mechanical treatment, by studying a spin-1/2 particle coupled to a finite-sized directional quantum reference frame, a spin-$l$ system, which models an external field. With this concrete model together with a bosonic thermal bath, we analyse the back-action a finite-size field suffers during a quantum-mechanical work extraction process, the effect this has on the extractable work, and highlight a range of assumptions commonly made when considering such processes. The well-known semi-classical treatment of work extraction from a pure qubit predicts a maximum extractable work $W = kT \log 2$ for a quasi-static process, which holds as a strict upper bound in the fully quantum mechanical case, and is only attained in the classical limit. We also address the problem of emergent local time-dependence in a joint system with globally fixed Hamiltonian.
\end{abstract}


\maketitle

\section{Introduction\label{sec:Introduction}}
The concept of work plays a crucial role in thermodynamics. It can be seen as a highly ordered form of energy, as opposed to the very disordered heat which (in macroscopic thermodynamics) is essentially random motion of particles. Yet in quantum physics the notions of order and disorder are very subtle. Hence it might not be too surprising that a generally accepted notion of work has not yet been established for quantum systems. Various considerations and proposals do exist (see e.g. \cite{Horodecki2011,Skrzypczyk2013,Brunner2012,Linden2010a,Linden2010,Alicki1979,Alicki2004}), but no consensus has been reached, and it is not inconceivable that multiple complementary notions of work exist at the quantum scale, as is the case for other quantities such as the free energy \cite{Horodecki2011}.\\
In this article we shall attempt to further bridge the gap between classical and quantum thermodynamics, by generalising a well known semi-classical model of work extraction to a fully quantum-mechanical model. By doing so, we attempt to narrow in on a more operational definition of work at the quantum scale, highlight conceptual and technical obstacles and pave the way for further investigations.

\subsection{The Dogma}
\begin{figure}[h]
\begin{center}
\includegraphics[width = \columnwidth]{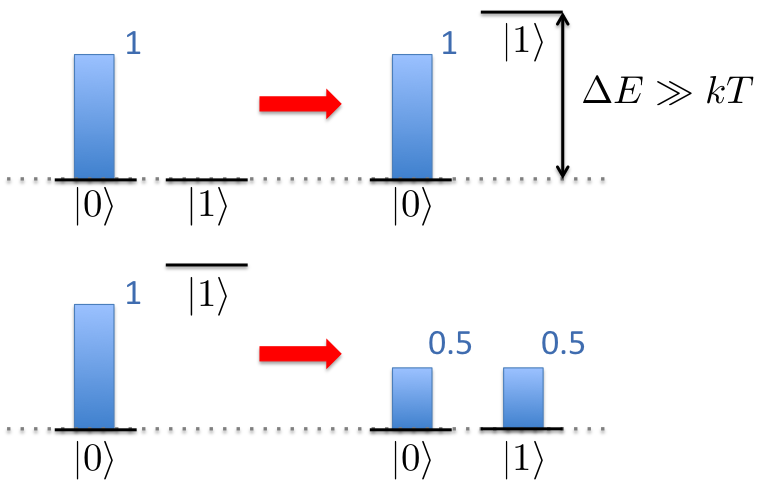}
\caption{(Color online) Top: The qubit starts in a known pure state $\ket{0}$. The unpopulated level $\ket{1}$ is then raised to a high energy without any work cost.\\
Bottom: The qubit is now coupled to a thermal bath and allowed to equilibrate. The raised level is then gradually lowered back to its originally position while keeping the qubit in equilibrium with the bath. Eventually the qubit is in a completely mixed state, with an increase in (von-Neumann) entropy of $\Delta S = \log 2$, and an amount of work $W= kT\log 2$ is said to be extracted during the process.}
\label{Dogma}
\end{center}
\end{figure}

Given a qubit in a known pure state, say the state $\ket{0}$, it is possible to convert the information about the state into work. The protocol that is generally quoted (see e.g. \citep{Alicki2004,Renner2011}) employs an external classical field, usually a magnetic field in the case of spin-qubits, which is gradually coupled to the system in order to raise the unoccupied state to a high (or ideally infinite) energy $E$. This raising process is schematically depicted in the top half of Figure \ref{Dogma}.
\\
After raising the unoccupied level, the qubit is coupled to a thermal bath at temperature $T$ and allowed to equilibrate. If the level was raised to a sufficiently high energy in the previous step, such that $E\gg kT$, this equilibration is a quasi-static process\footnotetext{$k$ is the Boltzmann constant. In the remainder of this article we shall use $kT$ and $\beta^{-1}$ interchangeably.}. The level is then lowered back to zero, slowly enough so that the system stays in thermal equilibrium with the bath throughout the entire lowering process, i.e. the qubit undergoes a quasi-static, isothermal transformation. The final state of the qubit is maximally mixed, with equal population in both levels, corresponding to a change in (von-Neumann) entropy of $\Delta S =\log 2$. Applying the first and second law of thermodynamics \cite{Fermi1956}, the work obtained in the process is thus said to be 
\begin{equation}\label{eq:kTlog2}
W=kT \log 2.
\end{equation} 
Running all the processes in reverse gives the well-known Landauer erasure protocol \cite{Landauer1961,Bennett2003,Faist2012,Hilt2011,Plenio2001}, with cost of erasure $W=kT \log 2$.\\ 
This protocol is almost universally accepted, yet it raises some very obvious questions. Particularly, what actually happens to the extracted work? And what if the field is not classical and of infinite strength but also a finite-sized quantum system which can evolve and develop correlations with the qubit? The first question is usually answered by saying that the field gains in its ability to do work, whereas the second question has to our knowledge not been seriously addressed in the literature. Another question that arises is what it even means for a classical field to gain in its ability to do mechanical work, since one could easily argue that a classical field already has an infinite ability to do work by simply driving Rabi oscillations on an arbitrary number of qubits and raising them to their excited state. We want to address these questions in a fully quantum-mechanical framework.\\

\subsection{Outline}
This article is structured as follows. In section \ref{sec:Model} we give a brief overview of the constituents of our model and the mathematical structure underlying them. In section \ref{sec:TimeDep} we study this model using a specific Hamiltonian, with a time dependent coupling between the system and field. We study the amount of extractable work and the back-action on the field, and compare the results with the semi-classical protocol. We show that the explicit time dependence will raise some questions about external control. Section \ref{sec:TimeIndep} will then look at an alternative coupling-Hamiltonian without any explicit time-dependence and a minimum amount of external control. In section \ref{sec:Bosonic} we depart from the idealised notion of the bath and introduce a more realistic bosonic bath into the time-independent model. Finally, in section \ref{sec:Weight} we add a potential work storage system in the form of a quantum weight.\\

\section{The Model\label{sec:Model}}
\begin{figure}[t]
\begin{center}
\includegraphics[width = \columnwidth]{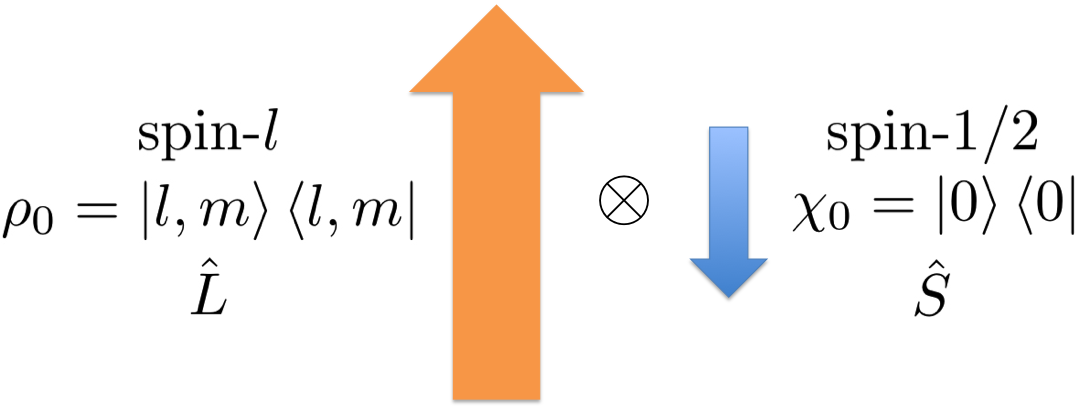}
\caption{(Color online) Schematic of the joint system, consisting of a \mbox{spin-$l$} reference frame and a spin-$1/2$ qubit, associated with the angular momentum operators $\hat{L}$ and $\hat{S}$ respectively, in the particular initial state $\sigma_0 = \ket{l,m}\bra{l,m}\otimes\ket{0}\bra{0}$.}
\label{Systems}
\end{center}
\end{figure}

The semi-classical model described in section \ref{sec:Introduction} consists of three very basic building blocks. At the heart of the protocol is the qubit, initially starting in a know pure state. This qubit is then coupled to two additional systems, the first being the classical external field, and the second the thermal bath at temperature $T$.\\
In our model we retain the qubit as in the original protocol\footnote{In the following we will use the words `qubit' and `system' interchangeably unless it is obvious that `system' refers to the joint system of quit plus external reference.}, realising it as a \mbox{spin-$1/2$} particle. We associate this spin-1/2 particle with the usual angular momentum operator $\hat{S}$, obeying the $SU(2)$ commutation relations, with components $\hat{S}_i$ proportional to the Pauli matrices $\hat{\sigma}_i$. We further assume that the qubit is initially in the state 
\begin{eqnarray}\label{eq:chi0}
\chi_0 &=& \ket{0}\bra{0} \nonumber \\
&=& \frac{1}{2}\bigl(\mathbb{1} - \hat{\sigma}_z\bigr),
\end{eqnarray}
the eigenstate of $\hat{S}_z$ with negative eigenvalue. We can think of this as the particle being fully polarised in the negative $z$-direction.\\
The classical field that provides the level splitting of the original model is replaced by a quantum system. In particular, we model the field by a spin-spin coupling to a directional quantum reference frame \cite{Bartlett2007,Poulin2007,Boileau2008,Angelo2011,Ahmadi2010,Bhaumik1975}, a spin-$l$ system\footnote{This system is referred to as the `field' or `reference', as it reproduces the coherent interactions and level-splitting of, for example, a spatially extended magnetic field.} described in as similar way as the qubit, with angular momentum operator $\hat{L}$. Even though our considerations allow for a more general treatment, we shall assume for simplicity that the reference starts in a state $\ket{l,m}$, the eigenstate of $\hat{L}_z$ with eigenvalue $m$, or in one of these states rotated around the $y$-axis. The generator of this rotation is $\hat{L}_y$, hence the most general initial state we consider for the reference can be written as
\begin{equation}\label{eq:rho0}
\rho_0 = e^{-i\phi\hat{L}_y}\ket{l,m}\bra{l,m}e^{i\phi\hat{L}_y},
\end{equation}
where $\phi$ denotes the angle of rotation with respect to the $y$-axis. We will come back to this general state in section \ref{sec:TimeIndep}, but for the preceding sections will assume the even simpler initial state with $\phi=0$. Figure \ref{Systems} schematically shows the qubit and reference for exactly this case of $\phi=0$. We shall refer to the combined initial state as
\begin{equation}\label{eq:sigma0}
\sigma_0 = \rho_0 \otimes \chi_0.
\end{equation}
\\
The third ingredient, the bath, is modelled in two different ways. In sections \ref{sec:TimeDep} and \ref{sec:TimeIndep} we shall consider the bath simply as a black box system which can instantaneously replace another system's state with its respective Gibbs state
\begin{equation}\label{eq:gibbs}
\tau = \frac{e^{-\beta H}}{\mathcal{Z}},
\end{equation}
where
\begin{equation}\label{eq:par}
\mathcal{Z} = \rm{tr}\bigl[e^{-\beta H}\bigr],
\end{equation}
is the partition function, $H$ is the Hamiltonian of the system under consideration, and $\beta=1/kT$ is the bath's inverse temperature. This version of the bath is equivalent to the one considered in the semi-classical protocol. It is a reasonable approximation to the actual situation as long as the actual thermalisation timescale is very short compared to the system dynamics, and the spectrum of the bath is sufficiently broad.\\
This model is still very idealised though, and depends on various assumptions such as infinitely strong coupling (or infinitely slow processes). From section \ref{sec:Bosonic} onwards we shall depart from this idealised notion and replace it with a more realistic bosonic bath coupled to the qubit. One might be tempted to do so straight away by simply deriving and solving a Markovian master equation, but in section \ref{sec:Bosonic} we show that it is not as straight forward as one might expect, due to the fact that the bath only interacts with a subsystem of a larger joint system. The Markovian master equation approach seems to be unable to make this distinction. Hence we defer discussion of a more realistic bath until section \ref{sec:Bosonic} and start with the simplified bath model outlined above to first provide some intuition of the systems behaviour.

\section{Time-Dependent Hamiltonian\label{sec:TimeDep}}
In the absence of any coupling the qubit's (and the reference's) energy levels are fully degenerate. However, if we introduce a coupling between the two systems we can break the degeneracy. Specifically, we consider a time-dependent coupling Hamiltonian of the form 
\begin{equation}\label{eq:Ham1}
H(t) = f(t) \hat{L}\cdot\hat{S},
\end{equation}
where $f(t)$ is some tuneable coupling strength, which we assume to be zero initially, $f(0)=0$.\\ 
Since there is initially no population in the state $\ket{1}$ of the qubit (c.f. equation (\ref{eq:chi0})), we can freely raise or lower this state without any energy cost. We do so by turning on the coupling to the reference (see Appendix \ref{app:Raising} for more details). Wherever we present any explicit calculations in this section, we shall assume the coupling to be linear in time for simplicity, 
\begin{equation}\label{eq:f}
f(t) = C\pm\frac{t}{l}
\end{equation} 
where the $1/l$ factor was chosen to make the level splitting induced by different size references comparable and $C$ is some constant offset. The reference frame is assumed to be initially in a state of the form given by (\ref{eq:rho0}).\\
Once we have the states split, say we wait for $t=1$, we couple the qubit to a heat bath at temperature $T$ and thermalise it. As mentioned in section \ref{sec:Model} this thermalisation process is assumed to occur instantaneously.\\
From this time on we keep the system coupled to the bath, and slowly (i.e. quasi-statically) tune down the coupling to the reference to extract work from the system. We consider this process in infinitesimal steps, each consisting of a thermalisation of the qubit followed by a change of the Hamiltonian and an infinitesimal joint evolution, and finally integrate over all the steps to get the full evolution. We may thus study what the back-action on the reference frame is and how much work we can extract.

\subsection{The Protocol}
Given the Hamiltonian in eq. (\ref{eq:Ham1}) we can look at the energy levels of the qubit. In order to do so we define a reduced Hamiltonian, which is found by taking the product of the total Hamiltonian and the other subsystems state tensored with the identity, and tracing over the other subsystem. We justify this definition in section \ref{sec:Bosonic}. Hence the qubit's reduced Hamiltonian is given by
\begin{eqnarray}\label{eq:Ham1Qu}
H_s(t) &=& \rm{tr}_r\bigl[(\rho(t)\otimes\mathbb{1})H(t)\bigr] \nonumber \\
&=& f(t) \braket{\hat{L}}\cdot\hat{S},
\end{eqnarray}
where $\rm{tr}_r[...]$ refers to a trace over the reference's Hilbert space. This Hamiltonian has two eigenvalues, $E_+(t)$ and $E_-(t)$. In order to make our model conceptually as close as possible to the original protocol of section \ref{sec:Introduction}, we want to fix the state $\ket{0}$ during the evolution. To do so we can add an additional term of the form $g(t)\mathbb{1}$ to the Hamiltonian, specifically
\begin{equation}\label{eq:Ham2}
H(t) = f(t) \hat{L}\cdot\hat{S} - E_-(t)\mathbb{1},
\end{equation}
which gives us
\begin{eqnarray}\label{eq:levels}
E_0(t) &=& 0 \nonumber \\
E_1(t) &=& E_+(t) - E_-(t),
\end{eqnarray}
as the energies of the states $\ket{0}$ and $\ket{1}$ respectively. This addition of a constant energy offset does not alter the actual physics, such as the evolution of the systems or the extractable work. In this sense, the Hamiltonians (\ref{eq:Ham1}) and (\ref{eq:Ham2}) can be considered equivalent.\\ 
As a specific example, consider the case where $\braket{L_x} = \braket{L_y} = 0$. Then from (\ref{eq:Ham2}) and (\ref{eq:levels}) we find for the system's reduced Hamiltonian and corresponding energy levels
\begin{equation}\label{eq:Ham2Qu}
H_s(t) = f(t) \braket{\hat{L}_z}\hat{S}_z + \frac{1}{2}f(t) \braket{\hat{L}_z}\mathbb{1}
\end{equation}
and
\begin{eqnarray}\label{eq:levels2}
E_0(t) &=& 0 \nonumber \\
E_1(t) &=& f(t)\braket{L_z}.
\end{eqnarray}
Since $E_0(t) = 0$ $\forall$ $t$ we shall drop the subscript $1$ from $E_1(t)$ and simply refer to it as the energy splitting $E(t)$ henceforth.\\ 
This shows that our Hamiltonian achieves the desired effect. Keeping the occupied $\ket{0}$ level at fixed zero energy, we can freely raise or lower the initially unoccupied $\ket{1}$ state either directly by varying the coupling strength $f(t)$, or indirectly by varying the reference's polarisation in the $z$-direction.\\ 
Having covered the basic ideas related to this particular Hamiltonian, we now take a closer look at the evolution of the joint system during the work extraction process. In appendix \ref{app:Raising} we show that the initial part of the protocol, the raising of the unoccupied level, can be made trivial so that the joint system remains in the state $\sigma_0$.\\
The lowering process is less straightforward to analyse. At $t=\mathcal{T}$, with the state $\ket{1}$ raised to an energy $E(\mathcal{T})$ given by (\ref{eq:levels}), we couple the qubit to our thermal bath at inverse temperature $\beta$ and gradually reduce the coupling $f(t)$ between qubit and reference back to zero. We do this in infinitesimal steps of duration $dt$, each consisting of a thermalisation of the qubit, which is assumed to happen instantaneously, followed by a joint evolution of system and reference at fixed Hamiltonian, for a time $dt$. We consider again a total transition time $\mathcal{T}$ and a coupling function $f(t) = (\mathcal{T}-t)/l$, where for convenience we have reset $t=0$ to coincide with the beginning of the lowering process.\\
We now want to take a look at the thermal state of the qubit. From (\ref{eq:Ham2}) we find for the system's reduced Hamiltonian
\begin{equation}\label{eq:HamRed}
H_s(t) = f(t) \braket{\hat{L}}\cdot\hat{S} - E_-(t)\mathbb{1}.
\end{equation}
When coupling the qubit to the bath and thermalising it at time $t$ we thus find, according to (\ref{eq:gibbs}), the system to be in the Gibbs state
\begin{equation}\label{eq:thermalQ}
\chi(t) = \frac{1}{{\cal{Z}}(t)}e^{-\beta H_s(t)} = p_0(t)\ket{\tilde{0}}\bra{\tilde{0}}+p_1(t)\ket{\tilde{1}}\bra{\tilde{1}},
\end{equation}
where ${\cal{Z}}(t)= \rm{tr}[e^{-\beta H_s(t)}]$ is the partition function, $\ket{\tilde{0}}$ and $\ket{\tilde{1}}$ are the eigenstates of the reduced Hamiltonian (\ref{eq:HamRed}) with eigenvalues $0$ and $E(t)$ respectively, and 
\begin{eqnarray}
p_0(t) &=& \frac{1}{1+e^{-\beta E(t)}},  \nonumber \\
p_1(t) &=& 1-p_0(t).
\end{eqnarray}
Considering again, as above, the specific case for which $\braket{L_x} = \braket{L_y} = 0$, we find that $\ket{\tilde{0}} = \ket{0}$ and $\ket{\tilde{1}}=\ket{1}$ and
\begin{eqnarray}
p_0(t) &=& \frac{1}{1+e^{-\beta f(t) \braket{L_z}}} \nonumber \\
p_1(t) &=& \frac{e^{-\beta f(t) \braket{L_z}}}{1+e^{-\beta f(t) \braket{L_z}}}.
\end{eqnarray}
In the infinitesimal time step from $t$ to $t+dt$ the composite system evolves according to the unitary
\begin{equation}
dU(t) \equiv U(t,t+dt) = e^{-if(t) \hat{L}\cdot\hat{S} dt}.
\end{equation}
Using $\hat{L}\cdot\hat{S} = \frac{1}{2}[l\Pi_+-(l+1)\Pi_-]$ as shown in \cite{Ahmadi2010}, where $\Pi_{\pm}$ are projectors onto the $\ket{j=l\pm\frac{1}{2}}$ eigenspaces, we find, analogously to the expressions for the evolution operator during raising derived in Appendix \ref{app:Raising}, that 
\begin{equation}
dU(t)  = \Pi_+ + e^{-i d\gamma(t)}\Pi_-,
\end{equation}
where we have defined
\begin{equation}
d\gamma(t)  \equiv f(t) (l+\frac{1}{2}) dt.
\end{equation}

\subsection{Work Extraction}
\begin{figure}[t]
\begin{center}
\includegraphics[width = \columnwidth]{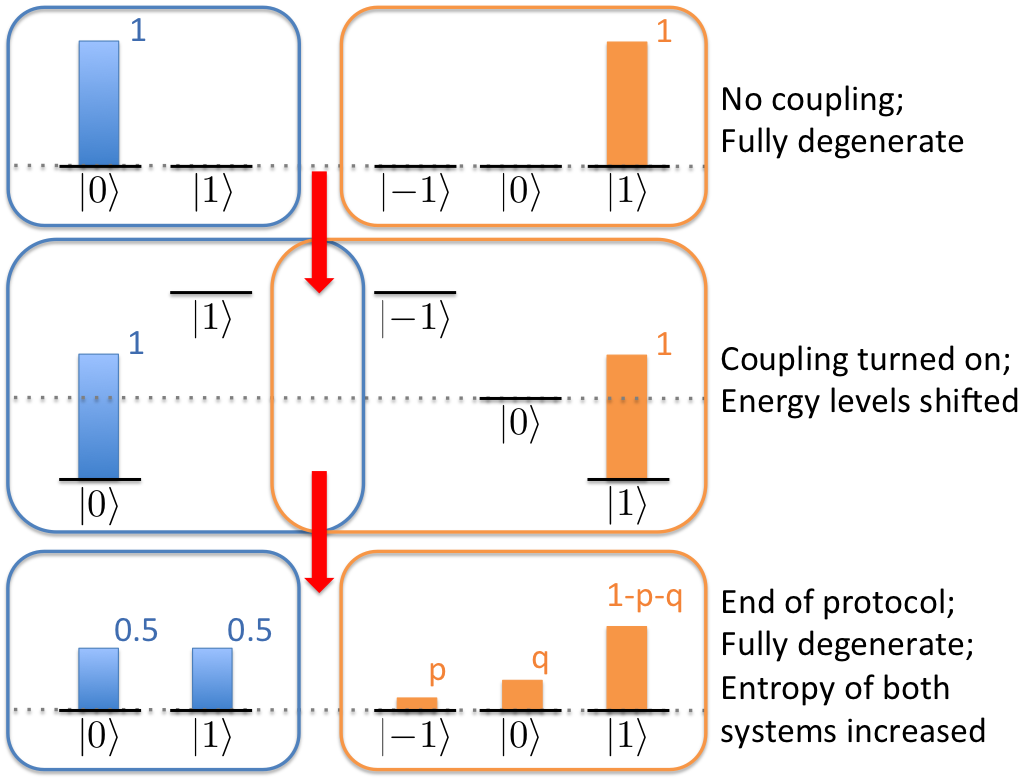}
\caption{(Color online) Schematic of the time-dependent Hamiltonian protocol using the example of an $l=1$ reference starting in the fully coherent state $\ket{1,1}$. Initially, without the coupling, $f(0)=0$, both the qubit's and the reference's states are fully degenerate. When the interaction strength is turned up, the degeneracy is broken as the energy levels shift. Once the maximum desired coupling is reached the qubit is couple to a thermal bath and the qubit-reference interaction $f(t)$ is slowly tuned back to zero again. As in the semi-classical protocol the qubit is maximally mixed after the protocol is finished. But in addition, the population of the reference has also shifted and the reference is not in a pure state anymore, having lost some of its polarisation and gained in entopy.}
\label{l1}
\end{center}
\end{figure}

\begin{figure}[tb]
\begin{center}
\includegraphics[width = \columnwidth]{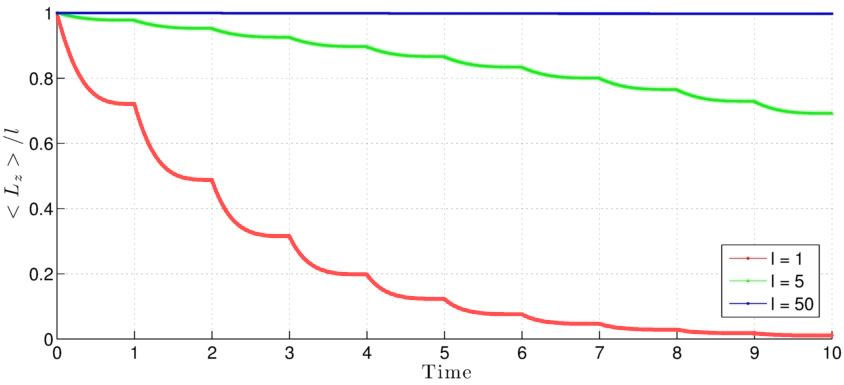}
\caption{(Color online) Scaled expectation values of the reference's $z$-component of angular momentum as a function of time for the same process as shown in Figure \ref{Work}. One unit of time is chosen to correspond to one iteration (ignoring the time required for raising the unoccupied level), i.e. this plot corresponds to the first ten iterations in Figure \ref{Work}. We observe a direct correlation between the value of $\braket{\hat{L}_z}$ and the extracted work in the corresponding iteration, but also between the rate of polarisation loss of the reference and a surplus in the extracted work beyond the semi-classical result.}
\label{AngMom}
\end{center}
\end{figure}

\begin{figure}[tb]
\begin{center}
\includegraphics[width = \columnwidth]{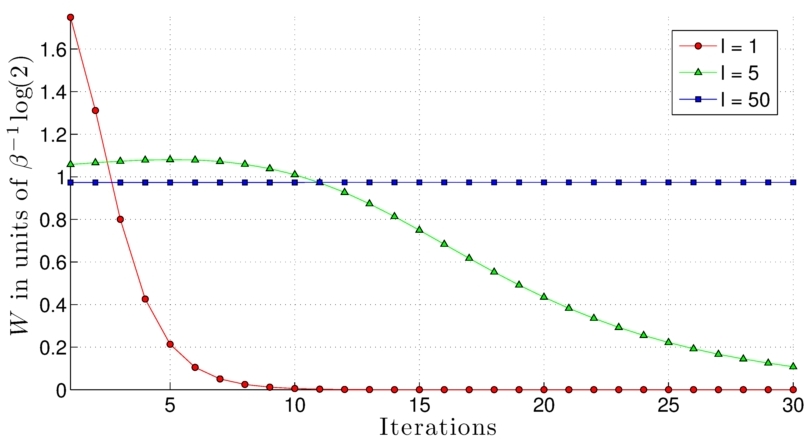}
\caption{(Color online) Work extracted from the joint qubit-reference-system over a single iteration, plotted against the number of iterations. The same reference is used for consecutive iterations, but with a new pure qubit. This plot clearly shows the tradeoff between single-shot work extraction and reference degradation. For small $l$ a large amount of work is initially extracted, but the reference rapidly loses its ability to induce a level splitting and extract further work (c.f. Figure \ref{AngMom}). For large $l$ the reference is nearly unperturbed and we approach the semi-classical $W= kT\log 2$ result, repeatable over many iterations with the same reference.}
\label{Work}
\end{center}
\end{figure}

At the core of all our considerations is the question of how much work we can extract during the entire process. At this point, following the common convention (see e.g. \cite{Gemmer2010,Alicki2004,Aberg2011}), we simply define the work extracted from the joint qubit-reference system in the infinitesimal time step $dt$ as
\begin{equation}\label{eq:work}
dW(t) \equiv \rm{tr}\bigl[\sigma(t) dH(t)\bigr].
\end{equation}
where $\sigma(t)$ is the state of the joint-system at time $t$ and $dH(t) = H(t) - H(t-dt)$ is the change in the Hamiltonian during the infinitesimal step $dt$. One of the main aims of the current investigation is to give work at the quantum scale a more operational meaning, and thereby test and justify (\ref{eq:work}), which forms the basis for many studies in quantum thermodynamics. We shall return to this thought from the end of section \ref{sec:TimeIndep} onward, but for the present considerations we simply accept (\ref{eq:work}) and use it as our definition of work.\\ 
The above steps give us all the ingredients to numerically study the work extraction process. Figure \ref{l1} shows the entire process (including the raising) schematically, analogous to Figure \ref{Dogma}, for the specific case of $l=1$ and the reference starting in the fully spin coherent state $\ket{1,1}$.
\\
One of the first observations from our numerical studies is that the qualitative features of the protocol seem to be entirely independent of the initial angle $\phi$ of the reference relative to the system (c.f. equation (\ref{eq:rho0})). Thus we shall in the following simply limit ourselves to the choice $\phi=0$ to simplify the problem. This observation is in line with our expectations, due to the rotationally invariant Hamiltonian (\ref{eq:Ham2}). When the system gets thermalised, it simply thermalises `along the direction' of the reference, independent of what exact direction in space this is. For $\phi=0$ we also find that neither system nor reference obtain any non-zero polarisation in the $x$- and $y$-directions. The problem is thus fully confined along a single axis.\\
Another immediate observation is that during the entire protocol, the polarisation of the reference frame, i.e. the value of $\braket{\hat{L}_z}$, appears to be monotonically decreasing, with a faster decrease for smaller $l$ as would be expected. Figure \ref{AngMom} shows $\braket{\hat{L}_z}/l$ for three representative sizes of $l$, over several iterations of the protocol, with the reference starting in the fully coherent state $\ket{l,l}\bra{l,l}$ before the first iteration\footnote{The simulations were done with $n=\mathcal{T}/dt=200$ discrete time steps for the lowering process and $\mathcal{T}\approx 5.013$. This very specific choice of $\mathcal{T}$ is explained in appendix \ref{app:Raising}. An inverse temperature $\beta=1$ was assumed for this and all other numerical results presented in this section.}. For each iteration we reuse the same reference, but introduce a new qubit in the pure state (\ref{eq:chi0}). We see that for small $l$ the reference loses its polarisation, and hence its ability to split the qubit's energy levels, very rapidly. For large $l$ such as $l=50$ in Figure \ref{AngMom} the reference barely experiences any backaction during the work extraction process and is almost unaffected for many iterations of the protocol. This is in good agreement with the intuition that $l\rightarrow\infty$ corresponds to a classical field as in the original protocol, which is unaffected by the process.
\\
In Figure \ref{Work} we show the work extracted from the joint qubit-reference system per iteration, as defined in (\ref{eq:work}), for the same values of $l$ and other parameters as used in Figure \ref{AngMom}. There are various aspects that are to be considered in this plot. The most obvious one being the fact that for small $l$ we appear to be extracting work $W > kT\log 2$ from the system during the first few iterations, breaking the limit set by the classical protocol. The simple explanation is in the degradation of the reference. In the original protocol the qubit turns into a maximally mixed state, extracting work at the expense of maximising its entropy. The same happens in our quantum protocol. But in addition to the qubit's entropy, the reference's entropy also increases as is schematically depicted in Figure \ref{l1}, with a more rapid increase for smaller reference sizes. This is reflected in the extracted work\footnote{In fact, if we only consider work extracted from the qubit, $dW_s(t) \equiv \rm{tr}\bigl[\chi(t) dH_s(t)\bigr]$, we find $W_s < kT\log 2$}. A small reference initially gives a large surplus in extractable work due to its rapid increase in entropy, but exactly this rapid entropy increase makes it useless as a reference very quickly. This is the second immediate observation, a direct tradeoff between initial work extraction and repeatability of the protocol. For very large references, which barely increase their entropy during one iteration due to the negligible backaction, the extracted work is barely changed from iteration to iteration and in the limit $l\rightarrow\infty$ we essentially recover the semi-classical result of $W = kT\log 2$, repeatable over an arbitrary number of iterations.
\\
We can also think of this surplus work extraction from the reference in a slightly different way. It seems fairly obvious that to create a highly asymmetric state such as the spin coherent state $\ket{l,l}\bra{l,l}$ we are assuming here, work has to be invested. It might be possible to think of this work as being locked/stored in the asymmetry of the state, and our protocol is able to release it again as extractable work. This is a very interesting notion since it seems to hint at a very close connection between different resource theories \cite{Brandao2011,Horodecki2012}, namely thermodynamics and asymmetry \cite{Lostaglio2014}. We appear to have a flow of asymmetry from the reference to the bath, analogous to heat flow in the case of energy considerations. Currently these thoughts are mostly speculations which need much more careful consideration, but at least point in interesting directions.\\
Looking solely at the work (\ref{eq:work}) extracted from the qubit we do not find any particularly surprising results. We find that in general it obeys both the classical limit $W < kT\log 2$, as well as the classical relation to the change in free energy. As expected, how close the work approaches $kT\log 2$ and how fast this value decreases over several iterations depends solely on the polarisation of the reference, i.e. on the amount of degeneracy breaking that can be induced in the qubit.

\subsection{The Issue of External Control}
Our main motivation in quantising the reference was to get a better picture of what exactly happens to the extracted work in the protocol. However, this question still remains open. In some sense it could be said that we have merely shifted the problem to the next level of abstraction. In the semi-classical protocol one could have said that the work is gained by the field. We have now explicitly quantised the field and can see that this is not the case for our reference model of the field, since the reference simply degrades and increases in entropy. However, in our model one could say that the work is transferred to the entity controlling the coupling strength $f(t)$.\\
We can think of what it physically means to change the coupling between the qubit and the reference. The simplest explanation is that the coupling $f(t)$ is simply a function of the distance between the two systems. Thus we can increase (decrease) $f(t)$ by moving the two systems closer together (further apart). Assuming our initial starting states, the total energy is actually lowered by turning on the coupling (see e.g. Figure \ref{l1}), thus there is a force pulling the systems towards each other. We could imagine attaching `weights' to our spins and using this force to lift the weights. Then as the qubit is being thermalised and the joint system's populations start to shift, less force is needed to reduce the coupling back to zero, i.e. less work needs to be done. Thus over the entire protocol this leads to a positive net gain in work.\\
These considerations give a nice intuitive feeling, but are not real explanations. We cannot simply attach weights to spins. In addition we have in some sense implicitly used equation (\ref{eq:work}), which we are trying to justify. The problem lies in the explicit time-dependence of the Hamiltonian (\ref{eq:Ham1}). It requires external control, some entity which we do not include in our quantum description. In the following section we introduce a way to avoid these issues and reduce any external control to a minimum by means of a time independent Hamiltonian.

\section{Time-Independent Hamiltonian\label{sec:TimeIndep}}
So far we have studied the work extraction from a single pure qubit coupled to a quantum-field modelled by a spin-$l$ directional reference frame. The two systems were coupled via a Hamiltonian of the form (\ref{eq:Ham1}). The explicitly time-dependent coupling is the issue of this model, since it requires external control. It does achieve quantisation of the external field to some extent, as opposed to the semi-classical model, but the question of what happens to the field is just replaced by the question of what happens to the entity that controls the parameter $f(t)$.\\
To minimise external control and the issues arising through it outlined in section \ref{sec:TimeDep}, we can consider a modification of our original model, free of any time-dependent parameters. Specifically, we shall look at the qubit-reference coupling Hamiltonian
\begin{equation}\label{eq:HamNew}
H =\sin\theta\hat{L}_z \otimes \hat{\sigma}_z+ \cos\theta\hat{L}_y\otimes \mathbb{1},
\end{equation}
where $0\leq\theta\leq\pi/2$ is a free parameter that allows for tuning the relative strength of the two terms, and $\hat{\sigma}_z$ is the Pauli $z$ matrix. The first term is the coupling between the qubit and the reference, where we have now assumed a $z$-axis bias of the coupling, as opposed to the rotationally invariant (\ref{eq:Ham1}) previously considered. The second term can be interpreted as the free Hamiltonian of the reference, with an energy associated to $y$-polarisation. This induces a precession of the reference around the $y$-axis which in turn leads to the desired evolution of the systems' states, without any external control. More precisely, under free evolution the reference's $y$-axis precession leads to a periodic increase and decrease in the value of $\braket{L_z}$, which due to the coupling given by the first term in (\ref{eq:HamNew}) induces a periodic raising and lowering of the qubits energy levels. The full evolution of the joint system is now encoded in the states of qubit and reference, with no hidden information such as the state of the entity controlling $f(t)$ in section \ref{sec:TimeDep}. 

\subsection{The Protocol}
As before, we start the protocol with the qubit in the known pure state $\chi_0 = \ket{0}\bra{0}$, i.e. fully polarised along the negative $z$-axis. The reference now starts in the initial state
\begin{equation}
\rho_0 = \hat{R}_y\bigl(\frac{3\pi}{2}\bigr)\ket{l,m}\bra{l,m}\hat{R}_y^{\dagger}\bigl(\frac{3\pi}{2}\bigr),
\end{equation}
where $\ket{l,m}$ is the $L_z$ eigenstate with eigenvalue $m$ and $\hat{R}_y(\frac{3\pi}{2}) = \exp{(-i\frac{3\pi}{2}\hat{L}_y)}$ is a rotation by $\frac{3\pi}{2}$ around the $y$-axis. This is exactly the form (\ref{eq:rho0}) with $\psi  = 3\pi/2$. Thus $\rho_0$ is an eigenstate of $L_x$ with eigenvalue $m$, pointing in the negative $x$ direction (assuming $m>0$). Note that due to the fact that the reference is polarised along $-x$ only, the value of $\braket{L_z}$ is initially zero which implies that the qubit's levels are completely degenerate, as should be the case at the beginning of the protocol. We again denote the joint system by $\sigma_0 = \rho_0\otimes\chi_0$.
The joint system now freely evolves for a time $t$ into the new state
\begin{equation}
\sigma(t) = U(t)\sigma_0U(t)^{\dagger},
\end{equation}
where
\begin{equation}\label{eq:U}
U(t) = e^{-iHt}
\end{equation}
and $H$ is the Hamiltonian (\ref{eq:HamNew}). However, we can note that the qubit is in an eigenstate of $\hat{\sigma}_z$ (with eigenvalue $-1$) and is thus invariant under the evolution (\ref{eq:U}). As long as the qubit remains in this state we can hence simplify the evolution of the joint state to 
\begin{equation} \label{eq:H_simp}
\sigma(t) = U_R(t)\rho_0U_R(t)^{\dagger} \otimes \chi_0,
\end{equation}
where
\begin{equation}\label{eq:UR}
U_R(t) = e^{-iH_Rt}
\end{equation}
and
\begin{equation}\label{eq:HR}
H_R =-\sin\theta\hat{L}_z  + \cos\theta\hat{L}_y.
\end{equation}
Taking a closer look at (\ref{eq:UR}) we see that it closely resembles the rotation operator
\begin{equation}\label{eq:D}
D(\hat{n},\phi) = e^{-i \phi \hat{n}\cdot\hat{L}}
\end{equation}
corresponding to a rotation around the axis defined by the unit vector $\hat{n}$ through an angle $\phi$ \cite{Sakurai2010}. In fact, we can rewrite (\ref{eq:HR}) as
\begin{equation}\label{eq:HRVec}
H_R = \hat{n} \cdot \hat{L}
\end{equation}
where we have defined
\begin{equation}\label{eq:nvec}
\hat{n} \equiv \begin{pmatrix} 0 \\ \cos\theta \\ -\sin\theta \end{pmatrix}.
\end{equation}
Substituting (\ref{eq:HRVec}) into (\ref{eq:UR}) and comparing it with (\ref{eq:D}) we see that the evolution of the reference is simply a rotation around an axis $\hat{n} = (0,\cos\theta,-\sin\theta)$ through an angle 
\begin{equation}\label{eq:phi}
\phi(t) = t \text{ mod } 2\pi.
\end{equation}
In other words, the reference experiences a rotation in space around the $+y$ and $-z$ axes, the amount of contribution of the two rotations being determined by the parameter $\theta$. As discussed above, the former leads to an increase in the reference's polarisation along the $z$-axis, which in turn induces a level splitting in the qubit, as can be seen from the qubit's reduced Hamiltonian
\begin{equation}\label{eq:HS}
H_S = \sin\theta\braket{\hat{L}_z}\hat{\sigma}_z + \cos\theta\braket{\hat{L}_y}\mathbb{1}.
\end{equation}
Note that the second term just leads to a constant energy offset, which is exactly large enough to keep the occupied $\ket{0}$ state fixed at zero energy. Also note that this reduced Hamiltonian (\ref{eq:HS}) is time-dependent due to the time dependence of the expectation values, despite the time independence of the total Hamiltonian.\\
We want to raise the unoccupied $\ket{1}$ state of the qubit as high as possible before starting to thermalise. We can thus ask again at which point in time this condition is satisfied. We shall again denote this time of maximal level splitting as $\mathcal{T}$, in line with the equivalent point in time in the protocol outlined in section \ref{sec:TimeDep}. As can be seen from (\ref{eq:HS}) the level splitting of the qubit is directly proportional to $\braket{\hat{L}_z}$, we thus define $\mathcal{T}$ as the time where $\braket{\hat{L}_z}$ reaches its (first) maximum. This occurs when the reference has rotated through an angle $\phi(\mathcal{T}) = \pi/2$ around $\hat{n}$. From (\ref{eq:phi}) it is immediately obvious that this happens at a time $\mathcal{T}=\pi/2$, for all values of $\theta$.
In addition, we can determine the value $\braket{\hat{L}_z}$ takes at this point. From (\ref{eq:nvec}) we see that the rotation axis $\vec{n}$ is at an angle $\theta$ measured from the $y$-axis to the negative $z$-axis. At time $\mathcal{T}$, $\braket{\hat{L}_z}$ thus reaches a maximum value of $\braket{\hat{L}_z}_{\mathcal{T}} = l\cos\theta$ where we have assumed that we are starting in a fully coherent state with $m=l$. From (\ref{eq:HS}) we see that this implies a maximum level splitting 
\begin{eqnarray}\label{eq:LMax}
E_\mathcal{T}&=&2l\sin\theta\cos\theta \nonumber \\
&=& l\sin(2\theta).
\end{eqnarray} 
A small value of $\theta$, $\theta<\pi/4$, implies that the rotation axis $\hat{n}$ given in (\ref{eq:nvec}) will be very close to the negative $z$-axis. This will lead to a smaller value of $\braket{\hat{L}_z}_{\mathcal{T}}$ and hence a smaller energy splitting $E_\mathcal{T}$ in the qubit. A large value $\theta>\pi/4$ on the other hand leads to an effective rotation axis $\hat{n}$ very close to the $y$-axis, which implies that the initially $x$-polarised reference will be almost fully rotated into the $z$-direction after quarter of a period, but the energy gap will still be a small factor in front of $\hat{\sigma}_z$. These considerations suggest that the ideal choice is $\theta = \pi/4$, giving both terms in the Hamiltonian equal weight.\\
(One might worry about the rotation breaking the assumption that the process is quasi-static. However, in our simple model we assume that the bath can thermalise the qubit instantly. In section \ref{sec:Bosonic} we consider a more realistic version of the bath and explore its effects on the process. However, for the numerics in the current section, we simply need to make our time steps $dt$ small enough in order to guarantee that the process is approximately quasi-static.)

\subsection{Work Extraction}
\begin{figure}[tb]
\begin{center}
\includegraphics[width = \columnwidth]{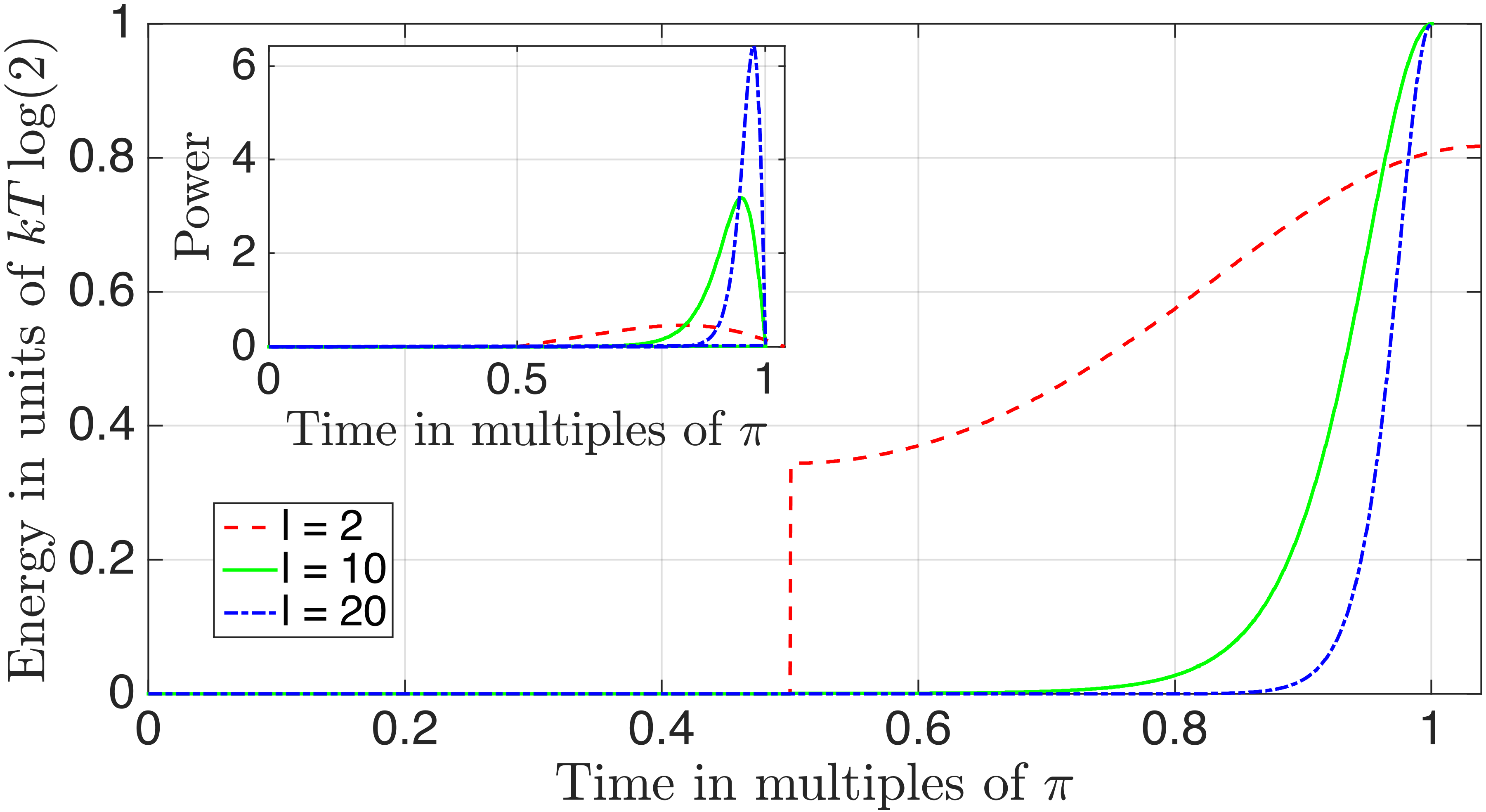}
\caption{(Color online) Energy stored in the reference as a function of time over one iteration of the protocol for $\theta = \pi/4$ and $dt=10^{-5}$ for different reference sizes. At $t=\pi/2$, $\braket{\hat{L}_z}$ reaches a maximum and the thermalisation/lowering process begins. We find that even with a comparatively low value of $l$ such as $l=10$ (green (middle we get very close to the semi-classical limit $W=kT\log 2$. In this specific case, we find that for $l=10$ an energy $E\approx0.9998kT\log 2$ is transferred from the bath into the reference, accompanied by and entropy gain of only $\Delta S \approx 7.9\times10^{-4}\log2$, resulting in $\Delta E -kT\Delta S \approx0.9990kT\log 2$. However, if the reference is too small, a considerably lower energy is transferred to the system. The discontinuous gap for $l=2$ is due to the idealised notion of an infinitely strong bath coupling. The inset shows the power as a function of time.}
\label{WorkVaryL}
\end{center}
\end{figure}

As before, at exactly this time $t=\mathcal{T}$ we couple the qubit to a thermal bath at inverse temperature $\beta$ and thermalise it, which again simply amounts to replacing its current state by the Gibbs state
\begin{equation} \label{eq:tauIndep}
\chi = \frac{e^{-\beta H_S}}{\mathcal{Z}},
\end{equation} 
where $H_S$ is the qubit's reduced Hamiltonian (\ref{eq:HS}) and ${\mathcal{Z}}$ is the corresponding partition function as defined in (\ref{eq:par}).\\
Analogously to the previous model, the protocol now proceeds in infinitesimal steps of duration $dt$, each consisting of an evolution of the joint system through the unitary $U(dt)$ as defined in (\ref{eq:U}), followed by a thermalisation of the qubit with the new reduced Hamiltonian. Note that now the evolution cannot be simplified as in equation (\ref{eq:H_simp}) since the qubit is no longer in an eigenstate of $\hat{\sigma}_z$. Taking $dt\rightarrow0$ gives us the quasi-static limit. The protocol is complete when $\braket{L_z}$ returns to a value of zero, at which point the energy levels of the qubit are degenerate again and the qubit is in a maximally mixed state.\\
This stage of the protocol is again less straight-forward to analyse than the raising process since the qubit's state is now constantly changing, which in turn leads to a constantly changing axis and angular velocity of the reference's rotation.
\begin{figure}[tb]
\begin{center}
\includegraphics[width = \columnwidth]{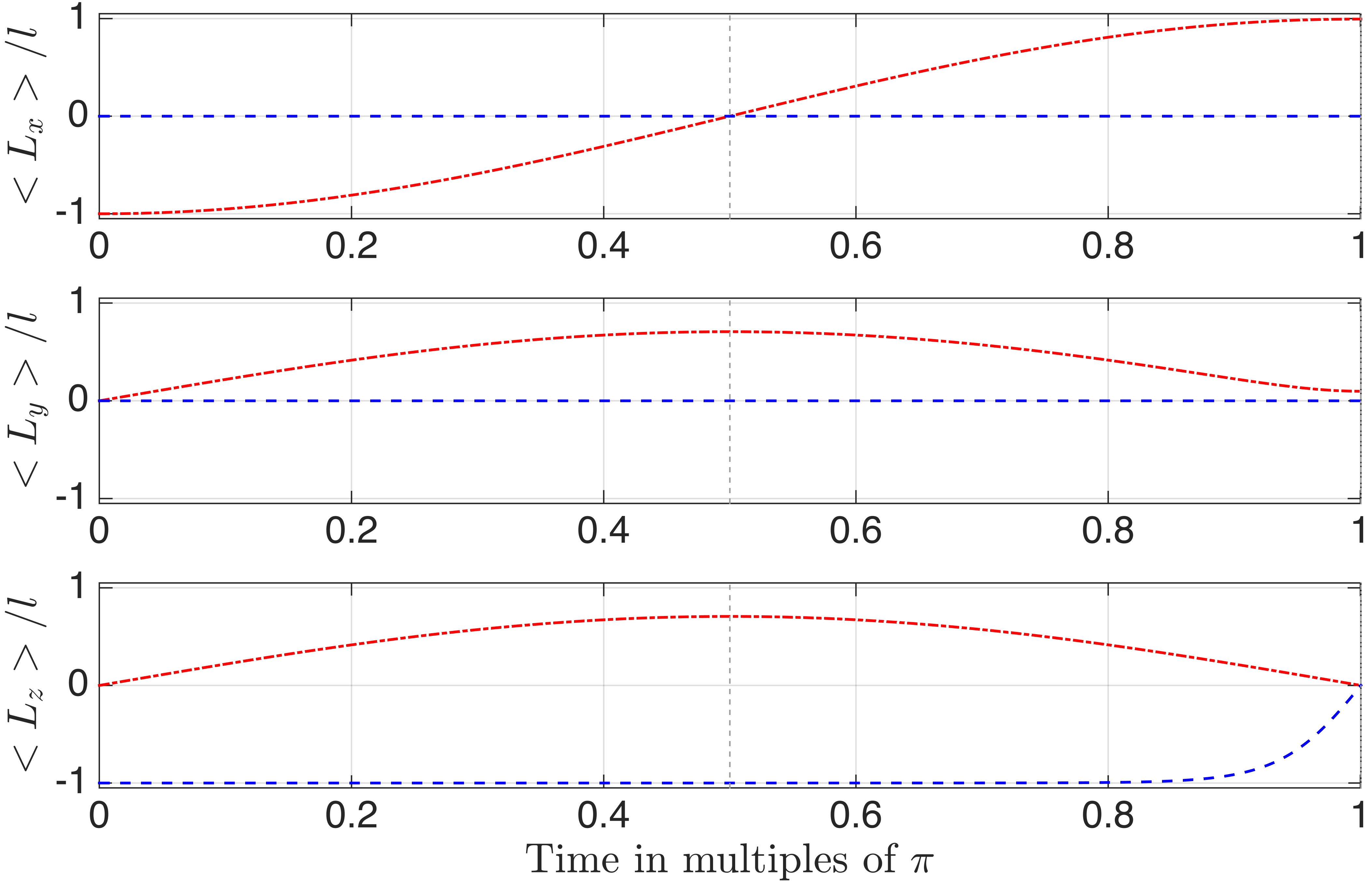}
\caption{(Color online) Scaled expectation values of the angular momentum components of the qubit (dashed blue) and reference (dash-dotted red) as a function of time for $l=10$,  $\theta = \pi/4$ and $dt=10^{-5}$. Initially at $t=0$ the reference starts fully polarised along $-x$ and the qubit along $-z$. Subsequently the reference rotates around an axis in the $y-z-$plane determined by the value of $\theta$ (c.f. Figure \ref{WorkVaryTheta}). At the time where $\braket{\hat{L}_z}$ reaches a maximum (vertical dashed line), the level splitting in the qubit is maximal and we begin thermalising it. This in turn affects the rotation of the reference, leaving it with some finite $\braket{\hat{L}_y}$ after completion of the protocol, i.e. when $\braket{\hat{L}_z} = 0$ again and the qubit is maximally mixed. By this mechanism, energy is transferred from the bath to the reference (c.f. Figure \ref{WorkVaryL}).}
\label{largeBExp}
\end{center}
\end{figure}\\
We now have all the basic building blocks to put together one complete iteration of our protocol. In the following we present results of a numerical analysis of the problem. Figure \ref{largeBExp} shows the scaled expectation values of the components of the angular momentum operators of both the qubit (blue) and an $l=10$ reference (red), for a single iteration of the protocol. We see that as expected the reference starts out fully polarised along $-x$ and then rotates around the axis $\hat{n}$ given in (\ref{eq:nvec}). For this particular choice of parameters, $\theta=\pi/4$, the rotation axis is exactly between the $y$- and negative $x-$axes. At the end of the raising process the reference has been rotated into the y-z-plane with $\braket{\hat{L}_z}_{\mathcal{T}} = \sqrt{2}l$ inducing a qubit energy gap of exactly $E_\mathcal{T}=l$ in accordance with (\ref{eq:LMax}).\\
As expected, the qubit stays fully polarised along $-z$ during the raising process. Once we start to thermalise the qubit at $t=\pi/2$ and the reference keeps rotating thus reducing the energy splitting again, the qubit gradually, approximately quasi-statically through many equilibrium states, approaches the maximally mixed state and the end of the protocol. Looking at Figure \ref{largeBExp} the reference appears almost unaffected by this. Yet, at the end of the protocol there is a small but finite polarisation along the $y$-direction remaining after $\braket{\hat{L}_z}$ has returned back to zero. We can also define a reduced Hamiltonian for the reference 
\begin{equation}\label{eq:HR2}
H_R =\sin\theta\braket{\hat{\sigma}_z}\hat{L}_z  + \cos\theta\hat{L}_y.
\end{equation} 
Thinking in terms of this reduced Hamiltonian helps us get a feeling for why a finite $\braket{\hat{L}_y}$ remains at the end of the protocol. As the magnitude of $\braket{\hat{\sigma}_z}$ decreases over the course of the protocol, the effective rotation axis tilts more and more towards the $y$-axis, until it coincide with it at the end of the protocol. We also see that this finite $\braket{\hat{L}_y}$ corresponds to a finite energy being stored in the reference. Figure \ref{WorkVaryL} shows the energy $E_R(t) = \rm{tr}[\rho(t) H_R(t)]$ stored in the reference over the full iteration.
\begin{figure}[tb]
\begin{center}
\includegraphics[width = \columnwidth]{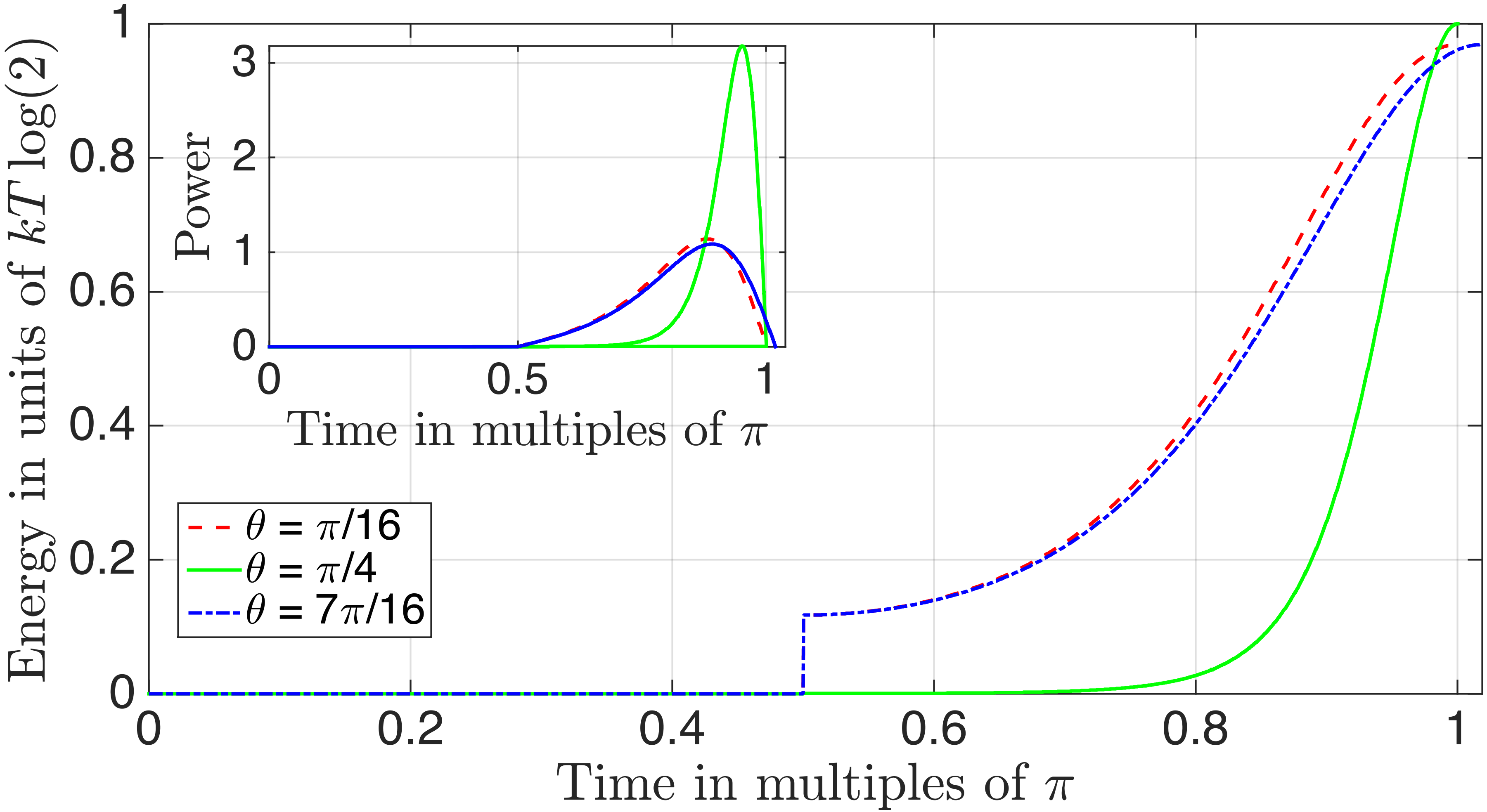}
\caption{(Color online) Same plot as Figure \ref{WorkVaryL} but for fixed $l=10$, $dt=10^{-5}$ and different values of $\theta$. We see that for $\theta=\pi/4$ we extract the maximum amount of energy. As $\theta$ tends to either zero or $\pi/2$ a lower amount of energy is transferred to the reference. The maximum level splitting is also considerably smaller for these values, as is evident from the discontinuity at the start of the thermalisation process.}
\label{WorkVaryTheta}
\end{center}
\end{figure}\\
In fact we see that at the end of the protocol an energy nearly approaching $kT\log 2$ has accumulated in the $l=10$ reference. This is particularly remarkable considering the small size of the reference (compared to the semi-classical case which essentially corresponds to an infinite size reference). We also see the the small $l=2$ reference experiences such a large perturbation during the thermalisation stage that its rotation period is noticeably prolonged. For completeness, Figure \ref{WorkVaryL} also shows the power $P_R=dE_R/dt$.\\
Again considering the reduced Hamiltonian (\ref{eq:HR2}) we see that at the end of the protocol, with $\braket{\hat{S}_z} = 0$, the energy stored in the reference is given by
\begin{equation}
E_R = \cos\theta\braket{\hat{L}_y}.
\end{equation} 
This shows that the larger we choose $\theta$, the more the reference must be polarised along $y$ after the protocol in order to store the same amount of energy. It also shows that if $\theta$ is too large for any given $l$ we cannot even in principle store a value of $kT\log 2$ in the reference. This consideration gives a maximum value for $\theta$
\begin{equation}
\theta < \cos^{-1}\Bigl(\frac{kT\log 2}{l}\Bigr).
\end{equation} 
This maximum value of $\theta$ is however no guarantee that we actually get close to pumping and energy $kT\log 2$ into the reference. Ideally we want $\theta$ to be as close to zero as possible in order to minimise the perturbation of the reference along the $y$-direction. However, a small $\theta$ also leads to a smaller energy gap and we have to be careful that the condition $E_{\mathcal{T}}\gg kT$ remains satisfied. These competing notions lead to an optimal value of $\theta=\pi/4$ as can be seen in Figure \ref{WorkVaryTheta}. Again we notice that the rotation period can be prolonged if the reference is perturbed considerably.\\
The gain in energy we observe also leads us back to the question of whether the energy gained by the reference can really be seen as work. We believe that this is the case to some extent, but that a part of the energy will also have to be classified as heat. If one were to apply the notion that work is an energy transfer which is not accompanied by entropy changes, and identify $W = \Delta E -kT\Delta S$ with the notion for work, the energies presented in Figure \ref{WorkVaryL} would almost entirely correspond to work, since for all three examples presented in the Figure the entropy gain is on the order of $8\times10^{-4}\log2$. However, so far these thoughts are merely speculation and a more detailed future analysis is necessary to settle these issues decisively. We have however shown that we are able to fully quantise all the constituents of the protocol and are thus able to account for all the energy flows without having to rely on equation (\ref{eq:work}). In addition our considerations are conceptually very similar to \cite{Skrzypczyk2013}, giving a very intuitive and operational picture of work at the quantum scale. We have also been able to avoid the issues of external control we uncovered in section \ref{sec:TimeDep}. The only external control which is required in the protocol (assuming one iteration) is to turn on the coupling between qubit and bath. We believe however that this is not an issue since the evolution leading up to this step is fully deterministic and we do not need to interact with the system in order to determine the time at which the coupling has to begin. There are no external entities which might gain or lose work in any process involved. Everything is accounted for in the states of our qubit and reference.\\
In appendix \ref{app:Iterations} we analyse the possibility of reusing one reference frame for multiple work extraction processes, as discussed for the time-dependent model in section \ref{sec:TimeDep}.

\section{Bosonic Bath\label{sec:Bosonic}}
\begin{figure}[tb]
\begin{center}
\includegraphics[width = \columnwidth]{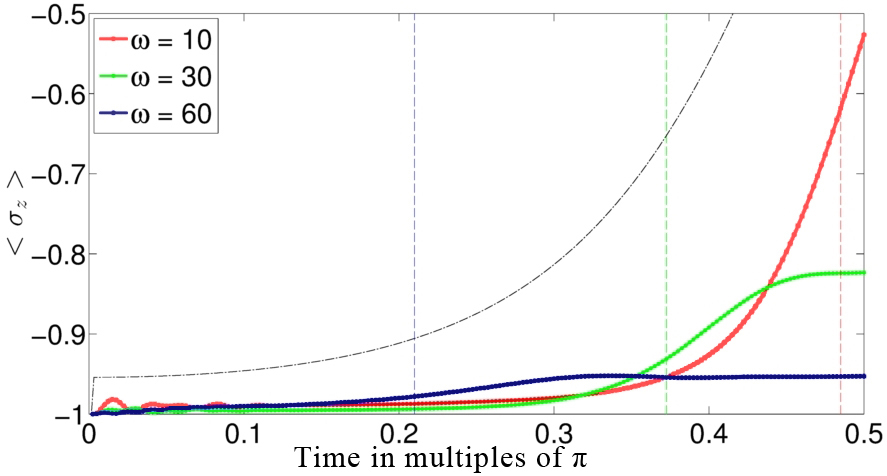}
\caption{(Color online) Expectation value $\braket{\sigma_z}$ for the qubit as a function of time, when coupled to a single-mode bosonic bath of frequency $\omega$. From right to left, the dashed vertical lines mark the times at which $\braket{L_z} = \frac{\omega}{\sqrt{2}}$ for $\omega=10$, $30$ and $60$ respectively. We see that these points exactly coincide with the times at which the qubit is most affected by the bath, i.e. they are in resonance. The dash-dotted black line gives $\braket{\sigma_z}$ as it would be obtained from using the idealised bath and reduced Hamiltonian presented in the previous considerations. The reference used in this plot is of size $l=75$, and the bath mode dimension is $D=7$. Other parameters are $\beta = 0.05$ and $\alpha = 2$.}
\label{SemiExact}
\end{center}
\end{figure}

The considerations above give us a good intuition for how the systems in our model generally behave, but, as alluded to previously, the idealised notion of the bath can lead to rather unphysical behaviour such as the discontinuities in Figures \ref{WorkVaryL} and \ref{WorkVaryTheta}. In this section we will replace the idealised bath with an actual bosonic bath coupled to the qubit, to add a further element of realism to the model. 
In order to do so we extend the Hilbert space of our model to $\mathcal{H} = \mathcal{H}_{\rm ref}\otimes\mathcal{H}_{\rm qubit}\otimes\mathcal{H}_{\rm bath}$ where $\mathcal{H}_{\rm bath}$ is the Hilbert space of the bath. Its free Hamiltonian is 
\begin{equation} \label{eq:HBath}
H_B = \mathbb{1}\otimes\mathbb{1}\otimes \sum_l \omega_l a_l^{\dagger} a_l,
\end{equation} 
where $a_l$ and $a_l^{\dagger}$ are the creation and annihilation operator of the $l$th mode with frequency $\omega_l$. In addition, the bath is coupled to the qubit via the interaction Hamiltonian
\begin{equation} \label{eq:HInt}
H_{\rm Int} = \mathbb{1}\otimes\sigma_x \otimes\sum_l g_l(a_l + a_l^{\dagger}).
\end{equation} 
The full Hamiltonian, where for simplicity we shall from here on assume $\theta = \pi/4$, is thus given by
\begin{eqnarray} \label{eq:HFull}
H =&\frac{1}{\sqrt{2}}&\Bigl[L_z\otimes\sigma_z + L_y\otimes \mathbb{1}\Bigr]\otimes\mathbb{1} \nonumber \\
&+&  \mathbb{1}\otimes\mathbb{1}\otimes \sum_l \omega_l a_l^{\dagger} a_l \nonumber \\
&+& \mathbb{1}\otimes\sigma_x\otimes\sum_l g_l(a_l + a_l^{\dagger}).
\end{eqnarray}
When presented with a Hamiltonian of this form the most obvious approach seems to be to derive a Markovian master equation and solve it. Our calculations however showed that this approach does not work in the present case where the bath is only coupled to one subsystem (the qubit) of a larger system. Despite the bath not being directly coupled to the reference, the Markovian approximation leads to the reference experiencing a direct thermalisation, and the entire qubit-reference joint-system simply thermalising to the Gibbs state given by the fixed level structure of the free Hamiltonian (\ref{eq:HamNew}). In fact, using the Markovian master equation approach, qualitatively this evolution is independent of the precise coupling between bath and joint-system. Even if we couple to the reference instead of the qubit (or a combination of both) by changing the final term in the Hamiltonian (\ref{eq:HInt}) to  e.g. $L_x\otimes\mathbb{1}\otimes\sum g_l(a_l + a_l^{\dagger})$ the behaviour qualitatively does not change. The Markovian master equation approach appears to be oblivious of the emergent local time-dependent structure of the individual subsystems, and only sees the temporally fixed level structure of the joint system, thus driving the entire system to the Gibbs state with respect to the full Hamiltonian, as opposed to driving the local system (in this case the qubit) towards thermal equilibrium, and only indirectly thermalising the remaining part of the joint system, as we would expect and as we explicitly show in the remainder of this section.\\ 
One might suggest that a possible approach to circumvent this issue is to decrease the qubit-bath coupling while decreasing the qubit-reference coupling. However, this leads to a change in the dynamics and never completely avoids the issue, only decreasing its severity. Instead we are able to provide a better way of understanding the thermalisation, based on resonance between qubit and bath-mode, which does not involve any approximation techniques.\\
If we want to thermalise only the qubit directly, with the precise thermalisation dependent on both the bath and the state of the reference, we need to approach the problem differently. The reason to use Markovian master equations in the first place is the impossibility of simulating the infinite number of field modes, each with an infinite number of states. To avoid the first problem, we simply assume that we are dealing with a single-mode bath, thus getting rid of one of the infinities. The other infinity can be reduced to a finite number by truncating the mode's local Hilbert space to the lowest $D$ energy levels, where $D$ is a finite integer. This approximation can be made arbitrarily accurate by choosing a larger $D$. The lower the temperature is relative to the bath's energy level spacing, giving by the mode frequency $\omega$, the smaller $D$ can be chosen. In addition, the interaction time between bath and system has to be limited. Both these restrictions assure that the highest levels never accumulate any non-negligible population and keep the bath close to being thermal at its characteristic inverse temperature $\beta$.\\
The simplest way to introduce a single mode bath into the model is to simply take a $D$-level system in the Gibbs state $\tau_0$ given by (\ref{eq:tauIndep}) with respect to the Hamiltonian $H_B=\omega a^{\dagger} a$, where $\omega$ is the frequency of the mode, couple it to the joint qubit-reference-system at the point where we want to start the thermalisation, and unitarily evolve the total system under the full Hamiltonian (\ref{eq:HFull}). In doing so, the bath will loose some of its `bathness', since it will evolve away from the original thermal state, but this will still give us some useful insights into the general behaviour of our new three-body system and guide us towards a more sophisticated treatment of the problem. In the context of a large thermal bath this corresponds to taking a (essentially negligibly) small sample of the bath which is 'wasted away'.\\
We shall denote the state of the new three-body system by $\sigma$, which is initially in $\sigma_0 = U_R(\pi/2)\rho_0 U_R(\pi/2)^{\dagger} \otimes \chi_0 \otimes \tau_0$, where $U_R$ is defined as in equation (\ref{eq:UR}) and the states $\rho_0$ and $\chi_0$ are (\ref{eq:rho0}) and (\ref{eq:chi0}) respectively. Subsequently the system evolves under the unitary $U(t) = \exp[-iHt]$ where  $H$ is the Hamiltonian (\ref{eq:HFull}), where the interaction part now consists of a single term 
\begin{equation} \label{eq:HIntSingle}
H_{\rm Int} = \mathbb{1}\otimes\sigma_x \otimes \alpha(a + a^{\dagger})
\end{equation} 
with coupling strength $\alpha$.\\
As the system evolves we again find that the reference undergoes a rotation very similar to what happened in the case of the idealised bath, thus gradually decreasing its $L_z$ expectation value to zero. The behaviour of the qubit is very interesting in this case. Its $\braket{\sigma_z}$ expectation value is plotted in Figure \ref{SemiExact} for a reference of size $l=75$ and baths of dimension $D=7$ with different frequencies\footnote{The bath dimension $D=7$ was chosen since for all $D\ge7$ no noticeable difference was found in the simulations.}. Despite being constantly coupled to the bath, it only `sees' the bath if the reference is in the correct state. More precisely, the qubit's state remains mostly unaffected, except around the time where $\braket{L_z} \approx \frac{\omega}{\sqrt{2}}$, where the qubit experiences an evolution towards a thermal state (\ref{eq:tauIndep}).
\\
Under closer examination we can see the importance of this result. In fact, it gives justification to the notion of reduced Hamiltonians that was used in the preceding sections of this paper. From equation (\ref{eq:HS}) the reduced Hamiltonian predicts that the qubit has a level splitting of $2\sin(\frac{\pi}{4})\braket{L_z}=\sqrt{2}\braket{L_z}$. Now we find that the qubit is affected most by the bath, when $\omega= \sqrt{2}\braket{L_z}$. Putting these two observation together implies that the qubit indeed has an effective level splitting of $\sqrt{2}\braket{L_z}$ and only interacts with the bath when it is resonant with it, i.e. when the level splitting roughly matches the bath's frequency $\omega$. In addition, its evolution towards a state that resembles the thermal state as would be given by the reduced Hamiltonian, is a further factor supporting the validity of the reduced Hamiltonian picture. These results are also somewhat reminiscent of the idea of virtual qubits presented in \cite{Brunner2012}, where the system only interacts with certain energy levels in the bath which are in resonance with the system. To summarise, we can say that the reduced Hamiltonian picture appears to be the correct description in the limit of a bath that instantaneously thermalises a system, which is exactly the assumption made in the previous sections.\\

\section{Lifting a Quantum Weight\label{sec:Weight}}
The above considerations all convincingly show how an amount of energy on the order of $kT\log 2$ is transferred to the reference system during the work extraction protocols. However, so far there is no convincing reason to believe that this energy can indeed be considered as work. To address this final gap in our considerations we introduce another new system, a quantum weight, similar to the approach in for example \cite{Skrzypczyk2013}.\\
The weight is initially not involved during the actual work extraction process described in the previous sections. Once this initial part of the protocol is finished we throw away the (approximately) maximally mixed qubit and couple the reference to the weight to try and convert the reference's excess in $\braket{L_y}$ it gained during the work extraction process into an unambiguous gain in mechanical energy, by raising the weight. The weight itself might simply be modelled by a particle in (one-dimensional) free fall with Hamiltonian
\begin{equation}\label{HWeight}
H_W = \frac{\hat{p}^2}{2\mu} + \mu g \hat{x}
\end{equation}
where $\hat{x}$ and $\hat{p}$ are the position and momentum operators, $\mu$ is the weight's mass, and $g$ the gravitational acceleration. This problem has been analysed in \cite{Voronin2006,Nesvizhevsky2000,Nesvizhevsky2002}.\\
In order to use the reference to lift the weight we need to introduce a coupling between the reference and the weight. We choose a Hamiltonian of the form
\begin{equation}\label{HWeightInt}
H_{RW} = \mathbb{1} \otimes \Bigl[\frac{\hat{p}^2}{2\mu} + \mu g \hat{x}\Bigr] + \cos\theta\hat{L}_y\otimes\mathbb{1} + \kappa \hat{A} \otimes \hat{p},
\end{equation}
where $\kappa$ is a tuneable coupling strength and $\hat{A}$ is some hermitian operator on the reference Hilbert space $\mathcal{H}_R$. The first and second terms of the Hamiltonian are the weight and the references' free Hamiltonians respectively, whereas the third term represents a coupling between the reference and the weight, which induces a translation of the weight dependent on the state of the reference.\\
One question is what operator $\hat{A}$ to choose to convert the excess energy stored into the $\braket{L_y}$ expectation value of the reference into positive translation of the weight. Finding an illuminating $\hat{A}$ is more awkward than might initially be assumed, and we do not construct one here, since it is more an engineering issue than one of fundamental physics. In particular, it amounts to pure mechanics, not thermodynamics, which is our main concern in this paper.\\
However, it is straightforward to see how the energy can be transferred from the reference to the weight. For conceptual simplicity one can approximate the continuous energy spectrum with a discrete one, and then the joint Hilbert space splits up into its energy subspaces $\mathcal{H}_E$ with constant energy $E$ as
\begin{equation}\label{HilbertSplit}
\mathcal{H}_R \otimes \mathcal{H}_W = \bigoplus_E \mathcal{H}_E.
\end{equation}
We can then define unitaries of the form
\begin{equation}\label{HilberSplit}
U = \bigoplus_E U_E,
\end{equation}
where $U_E$ acts on $\mathcal{H}_E$. Unitaries of this form are energy conserving and essentially represent swap operations between the reference and the weight, thus lifting the weight to higher energies while shifting the reference's population to lower energy levels. This approach is again reminiscent of the ideas proposed in \cite{Skrzypczyk2013}, and as in \cite{Skrzypczyk2013} and similar approaches, we have to be aware that while the weight will in general be lifted, it also experiences an energy spreading effect. The moral of this is that when using a large, but finite-sized system to hold extracted thermodynamic work we unavoidably face a probabilistic distribution over energies, and while there is a clearly a gain in usable mechanical energy, the `quality' of the energy requires further consideration. We contrast this average work extraction from the deterministic work extraction put forward in \cite{Horodecki2011,Brandao2013} for example.

\section{Conclusion\label{Conclusion}}
We set out on this research with the goal to get closer to understanding what it means to do work at the quantum scale. Our approach to doing so was to take a fresh look at an old protocol, the well-known semi-classical protocol of exacting an amount of work $W=kT\log 2$ from a pure qubit. This protocol has the weakness that not all its constituents are fully quantum mechanical, and what happens to the work is not accounted for in a quantum mechanical framework. Our model extends the original approach by replacing the semi-classical field with a finite-sized quantum reference frame, a spin-$l$ particle.\\
Using this new building block we looked at two specific incarnations of the model. The first of them uses a time-dependent Hamiltonian to couple the qubit and reference. Using this we were able to show that the reference (i.e. the external field) suffers from back-actions during the work extraction process, which lead to a degradation. These degradations have two distinct effects. On the one hand, they allow us to extract more work than the semi-classical $kT\log 2$ from the joint system (although the work extracted from the qubit remains less than $kT\log 2$, so this provides a useful illustration of a scenario that cheats the system). On the other hand, the degradation of the reference leads to a diminished ability to split the qubit's energy levels, which is crucial to extract further work. Hence there is a tradeoff between maximum single-shot work extraction and repeatability of the protocol. In general, as the size of the reference $l$ gets larger, we approach the semi-classical scenario as would be expected. The problem with this model is that, despite giving some new insights in the behaviour of a quantum mechanical field, does not answer the initial question. There is still an external entity receiving the work which is not accounted for in the quantum description.\\
To get rid of this external entity, we introduced a new coupling Hamiltonian which does not depend on time. In this model every flow of energy is fully encoded in the states of the qubit and the reference. It allowed us to show that (at least in this particular model) energy is indeed transferred onto the reference, i.e. the field, as is usually assumed in the semi-classical protocol. In addition, we were also able to recover the value $kT\log 2$ as an upper bound for this energy and show how the energy depends on the parameters of the model.\\
We added an additional element of realism to our model by introducing a bosonic bath coupled to the qubit, and using this were able to both pave the way to more sophisticated realisations of the model, as well as confirm the validity of previously made assumptions such as the reduced Hamiltonians.\\
Finally, we outlined a roadmap towards answering the question which part of the extracted energy can be considered work. In conclusion, we were able to uncover many subtleties that were previously either unknown or simply ignored, such as the issue of disordered energy flow into the field. The system we introduce still allows for many interesting questions, and promises to be a flexible, and tractable, model with which to study various fundamental aspects of quantum thermodynamics.

\section{Acknowledgments}
We acknowledge very useful discussions with Ahsan Nazir, Jake Iles-Smith, Mischa Woods and Matteo Lostaglio. This work was partially supported by the COST Action MP1209. TR supported by the Leverhulme Trust, MF supported by the EPSRC, and DJ supported by the Royal Society.

\bibliography{PureQubitWorkExtraction.bib}

\appendix

\section{Raising The Upper Level\label{app:Raising}}
We assume the system evolves freely under the Hamiltonian (\ref{eq:Ham2}) from an initial time $t=0$ to a final time $t=\mathcal{T}$. Thus, assuming $f(t)$ to be a monotonic function such as our particular choice in (\ref{eq:f}), the value of $\mathcal{T}$ determines the maximal level splitting. We can ignore the $- E_-(t)\mathbb{1}$ term  in (\ref{eq:Ham2}) since it does not contribute to the evolution and essentially use (\ref{eq:Ham1}) for our present considerations. We see that this Hamiltonian at different times commutes with itself, $[H(t),H(t')]=0$ $\forall$ $t,t'$. Thus, we find that the system evolves according to the unitary
\begin{equation}
U(\mathcal{T}) = exp\Bigl[-i \Phi(\mathcal{T}) \hat{L}\cdot\hat{S} \Bigr],
\end{equation}
where we have defined
\begin{equation}
\Phi(\mathcal{T}) \equiv \int_0^\mathcal{T} f(t) dt.
\end{equation}
Analogous to a very similar derivation presented in \cite{Ahmadi2010}, we were able to show that this can be written as
\begin{equation}\label{eq:URaising}
U(\mathcal{T}) = \Pi_+ + e^{-i\Gamma(\mathcal{T})}\Pi_-,
\end{equation}
where
\begin{equation}
\Gamma(\mathcal{T}) \equiv (l+\frac{1}{2}) \Phi(T)
\end{equation}
and $\Pi_{\pm}$ are the projectors onto the $\ket{j = l\pm\frac{1}{2}}$ subspaces of the joint Hilbert space \mbox{$\mathcal{H}_J = \mathcal{H}_L \otimes \mathcal{H}_S$}, where $\mathcal{H}_L$ and $\mathcal{H}_S$ are the Hilbert spaces of the reference and qubit, with dimension $d=2l+1$ and $2$ respectively.\\
In the main text we generally assume a linearly increasing coupling during the raising, of the form (\ref{eq:f}) with $C=0$, i.e. $f(t) = \frac{t}{l}$. In this case we get
\begin{equation}
\Phi(\mathcal{T}) = \frac{\mathcal{T}^2}{2l}
\end{equation}
and thus
\begin{equation}\label{eq:gammaT}
\Gamma(\mathcal{T}) = \frac{1}{4}(2+\frac{1}{l})\mathcal{T}^2.
\end{equation}
From (\ref{eq:URaising}) we see that if we choose $\Gamma(\mathcal{T}) = 2\pi n$ for any $n\in\mathbb{N}$, the evolution simply reduces to the identity and we can essentially ignore the entire raising part of the protocol, apart from noting that it induces the desired raising of the unpopulated state. In appendix \ref{app:Impossible} we take a closer look at the specific choice of $\mathcal{T}$.

\section{Impossibility of Perfect Work Extraction for the Time-Dependent Hamiltonian Model\label{app:Impossible}}
Most protocols that use a work extraction (or erasure) scheme which is based on splitting energy levels, say through an external (classical) field such as in the standard one presented in section \ref{sec:Introduction}, generally assume raising the state to infinite energy. This would correspond to $\mathcal{T}\rightarrow\infty$ in our protocol. It can generally be noted that $E\rightarrow\infty$ is an unphysical assumption, but with our specific model we are in a position to give a much more detailed and well founded study of this issue. Note that in this section we shall only consider work extracted from the qubit itself, ignoring the reference.\\
In general, again only assuming that $f(t)$ is monotonic, we find that the larger the value of $\mathcal{T}$, the quicker the reference frame degrades, leading to a diminished ability to extract further work. This makes intuitive sense, considering that larger $\mathcal{T}$ implies stronger coupling\footnote{Note that for our specific choice of $f(t)$ stronger coupling also implies longer coupling. If we demand our protocol to be approximately quasi-static, a correlation between maximal strength of coupling and duration of the protocol seems to be a sensible choice.} between the thermalised system and the reference, leading to quicker loss of asymmetry. However, if $\mathcal{T}$ is too small we do not raise the level particularly far, not satisfying the condition $E\gg kT$, and are unable to extract much work, giving a change in free energy $\Delta F$ (and hence by the inequality $dW \leq - dF$ also $W$) considerably lower than the optimal $\Delta F = kT\log 2$. We see that there is yet again a tradeoff between maximal work extraction in a single-shot scenario on one hand, and rapidly depleting the reference on the other. This raises the question about the optimal value of $\mathcal{T}$.\\
From the consideration in appendix \ref{app:Raising} we know that we require $\Gamma(\mathcal{T}) = 2\pi n $ to avoid non-trivial evolutions during the raising and keep the process as simple as possible. For our specific coupling function $f(t)=t/l$ this gives us from (\ref{eq:gammaT}) the first constraint on $\mathcal{T}$
\begin{equation}\label{eq:Tn}
\mathcal{T} = \sqrt{\frac{8\pi n}{2+l^{-1}}}.
\end{equation} 
Let us further assume that we want to have a potential change in free energy during the entire protocol no less than 
\begin{equation}
\Delta F = c [kT\log 2]
\end{equation} 
for some $0 < c < 1$. After some manipulations this gives us a second constraint on $\mathcal{T}$,
\begin{equation}
\mathcal{T} \geq - kT\log{[2^{1-c}-1]}.
\end{equation} 
To satisfy both conditions, we want $\mathcal{T}$ to be as given by (\ref{eq:Tn}), where $n$ is the smallest integer satisfying
\begin{equation}
n \geq \frac{2+l^{-1}}{8\pi \beta^2}\Bigl[\log{(2^{1-c}-1)}\Bigr]^2.
\end{equation} 
For large $l$ we can approximate these equations by 
\begin{equation}
\mathcal{T} \approx 2\sqrt{\pi n},
\end{equation} 
where $n$ is the smallest integer satisfying
\begin{equation}
n \gtrsim \frac{1}{4\pi \beta^2}\Bigl[\log{(2^{1-c}-1)}\Bigr]^2.
\end{equation}
For example if we want $\Delta F$ to be at least 99\% of its optimal value, i.e. $c=0.99$, we find that $n \gtrsim 1.96$, i.e. we choose $n=2$, giving $\mathcal{T}\approx 5.013$, which is the value of $\mathcal{T}$ we have used for the results in Figures \ref{AngMom} and \ref{Work}.\\
Generally we can conclude from these consideration of $\mathcal{T}$ that perfect work extraction (and in a similar way also perfect erasure) is impossible using a reference of finite size $l$, since it requires an infinite energy splitting, which immediately "destroys" the reference that creates this splitting. One has to accept a tradeoff between longevity of the reference, i.e. repeatability, and maximum work extraction. An interesting extension to this analysis might be to consider the cumulative work that can be extracted over a large/infinite number of iterations, and then find the ideal $\mathcal{T}$ which maximises this value for a given $l$.

\section{Multiple Iterations for the Time-Independent Hamiltonian Model\label{app:Iterations}}
Minimising the perturbation along the $y$-direction becomes particularly important if we want to reuse the reference for multiple iterations of the protocol. To keep extracting work, i.e. to keep increasing the reference's energy, we must be able to further increase $\braket{\hat{L}_y}$. However, for small $l$ we see that $\braket{\hat{L}_y}$ saturates quickly with the reference reaching a state fully polarised in the $y$-direction, at which point the protocol breaks down as we cannot extract any more work. In this short section we shall look at the idea of multiple iterations of the protocol in more detail.\\
There are three distinct ways in which we can repeat the protocol after finishing the first iteration. In all of them we first discard the previously used qubit which is now in a maximally mixed state. Removing this qubit does not affect the energy or state of the reference, hence this is a trivial process. The simplest way to proceed is to bring in a new pure qubit in a state similar to $\chi_0$ in (\ref{eq:chi0}), but polarised in the positive $z$-direction, the state $\ket{1}$ being the occupied level and $\ket{0}$ the unoccupied one. We assume that this can be done instantaneously (or at least fast relative to the evolution of the reference frame). The reference now keeps rotating, this time acquiring a negative polarisation $\braket{\hat{L}_z}$. Thus, by taking the qubit also polarised in the opposite direction we essentially return to the original situation. The protocol then follows the same steps as above until $\braket{\hat{L}_z}$ reaches zero again and the new qubit is in the maximally mixed state. We then iterate the process, using a qubit starting in $\ket{0}$ for every odd iteration, and one starting in $\ket{1}$ for even iterations.\\ 
The second method is mathematically essentially equivalent and leads to exactly the same results. Here we keep using qubits starting in the state $\ket{0}$, but keep flipping the Hamiltonian from $H$ to $-H$ after every iteration, reversing the references rotation. We shall use this method for our numerical analysis.\\
The third method also uses only qubits starting in the state $\ket{0}$. Here, after completing an iteration, we wait for the reference to freely evolve for an additional half period, at which time it is pointing in the original direction again (plus the perturbations it gained during the previous iterations). We then couple it to the new qubit. Again, this seems conceptually different from the other methods, but mathematically it amounts to exactly the same procedure. In the following we shall present results numerically obtained using the second method.\\
\begin{figure}[tb]
\begin{center}
\includegraphics[width = \columnwidth]{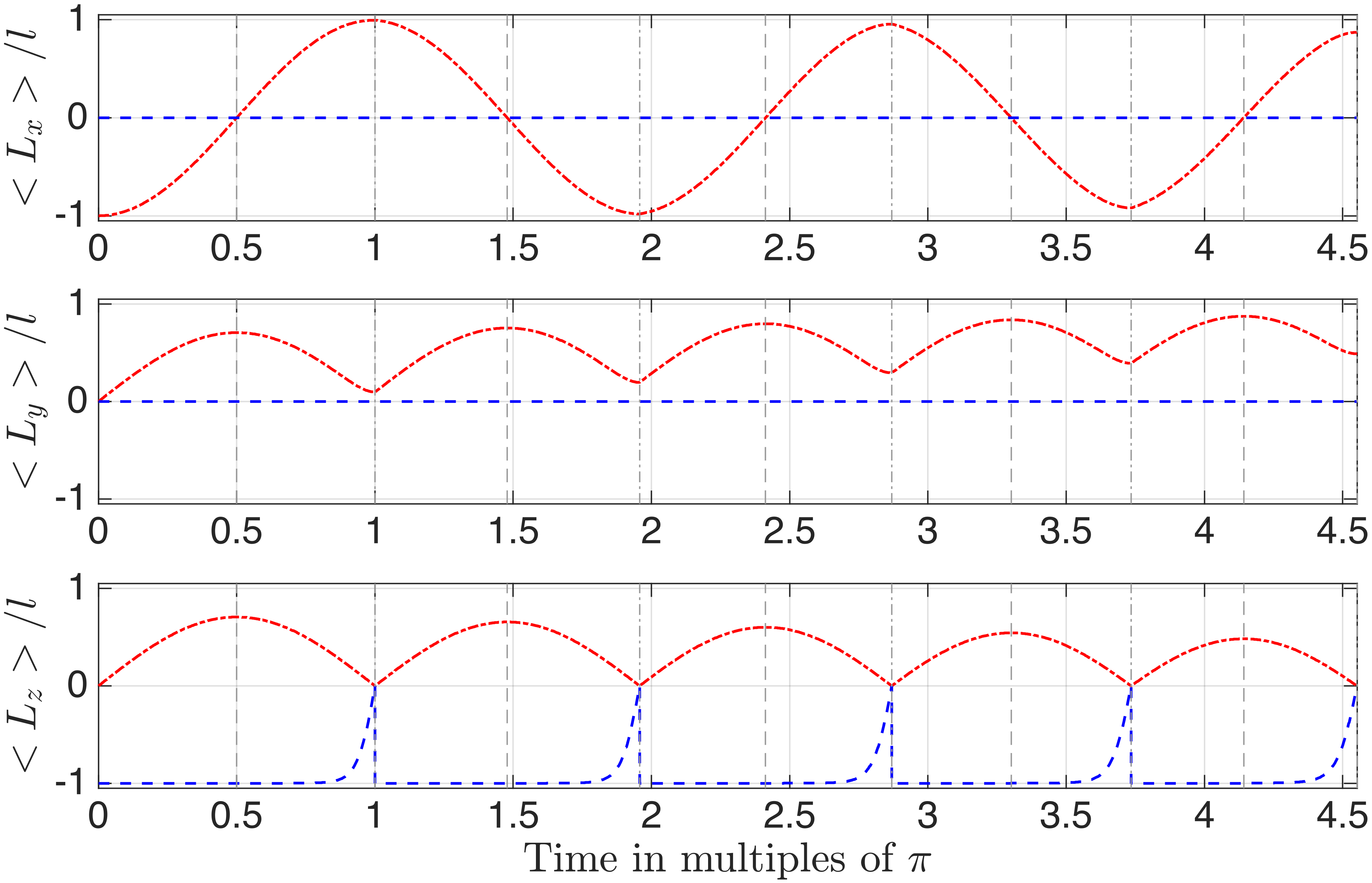}
\caption{(Color online) Scaled expectation values of the angular momentum components of the qubit (dashed blue) and reference (dash-dotted red) as a function of time for $l=10$,  $\theta = \pi/4$ and $dt=10^{-4}$. The same reference is reused for five iterations, each with a new pure qubit. A clear accumulation of the reference's polarisation in the $y$-direction is visible. The specific parameters were chosen to show a slow but visible degradation of the reference over the five iterations. Higher (lower) values of $l$ lead to slower (faster) degradation.}
\label{multExp}
\end{center}
\end{figure}
Figure \ref{multExp} shows the evolution of the expectation values of the angular momentum components over five iterations of the protocol for $l=10$ and $\theta=\pi/4$. As predicted, we see an accumulation of $\braket{\hat{L}_y}$ over consecutive iterations, which in the limit of many iterations completely saturates to $\braket{\hat{L}_y}\rightarrow l$ $\forall$ $t$. This is also reflected in Figure \ref{multEnt} which shows the energy of the reference as a function of time for the same process. 
\begin{figure}[b]
\begin{center}
\includegraphics[width = \columnwidth]{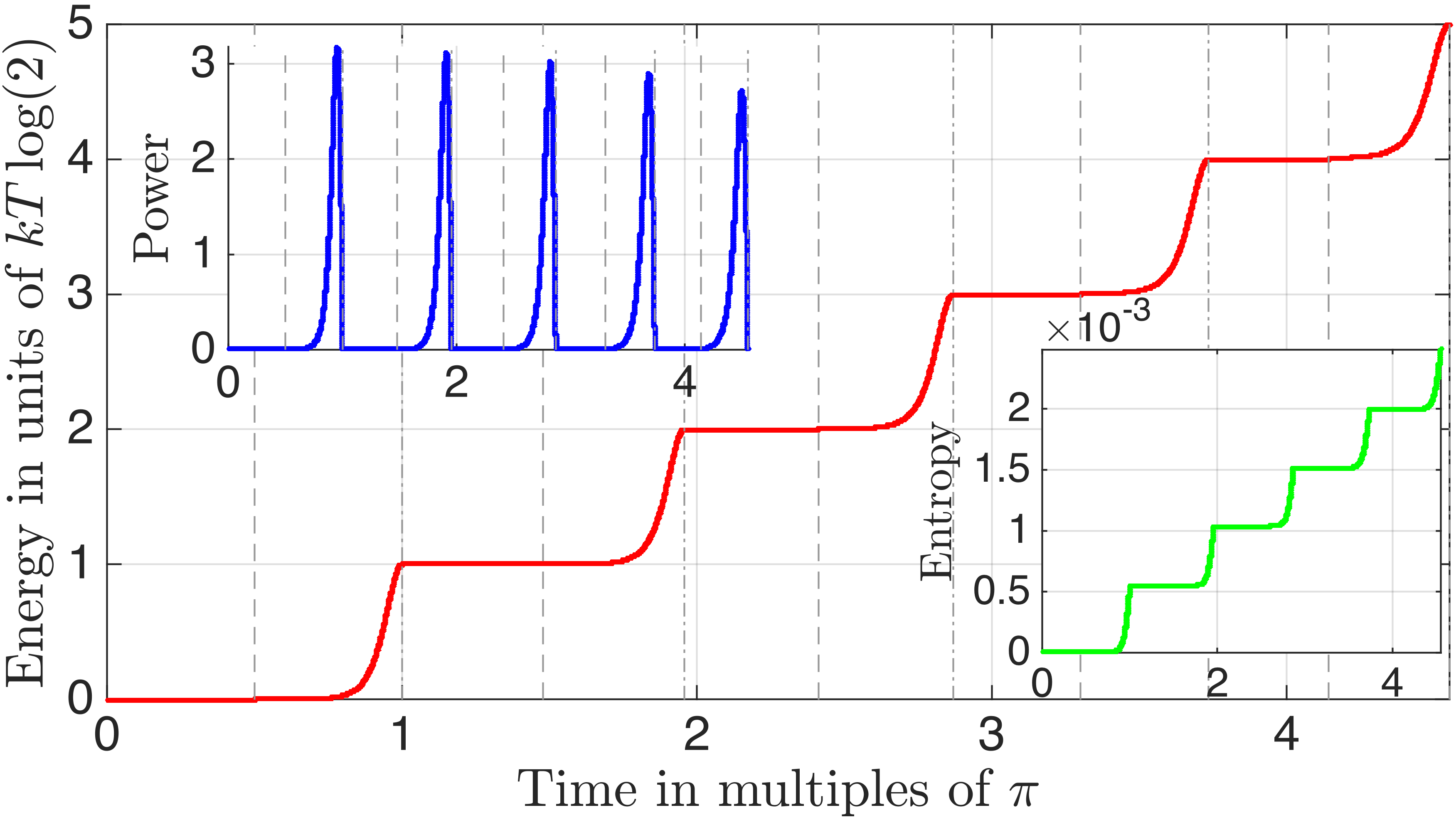}
\caption{(Color online) Energy stored in the reference as a function of time over five iterations of the protocol for $l=10$,  $\theta = \pi/4$ and $dt=10^{-4}$ as in Figure \ref{multExp}. During each consecutive iteration we transfer less and less energy onto the reference and the power drops. In the limit of infinitely many iterations, the reference completely polarises in the $y$-direction, saturating the amount of energy it can store. The green inset shows the (von-Neumann) entropy of the reference, which is monotonically increasing, also hinting at the constant degradation of the reference.}
\label{multEnt}
\end{center}
\end{figure}
We see that during the first iterations an energy of almost $kT\log 2$ is gained by the reference. However after a few iterations the energy gain starts to decrease. The plot of the power shows this effect  most clearly. In addition Figure \ref{multEnt} also contains a plot of the reference's von-Neumann entropy, which is monotonically increasing, while its purity (not shown) is constantly decreasing. Generally we find that the larger $l$, the more often we can iterate the process without seeing major signs of degradation of the reference. In the classical limit of $l \rightarrow \infty$ we find that we can repeat the protocol an arbitrary number of times without any degradation of the reference, as we expected.

\end{document}